\documentclass[%
 reprint, 
 superscriptaddress,
 amsmath,amssymb,
 aps,
floatfix,
longbibliography
]{revtex4-1}

\usepackage{graphicx}
\usepackage{bm}

\usepackage{xcolor}
\usepackage[colorlinks=true,citecolor=blue,linkcolor=magenta]{hyperref}

\newcommand{\ben}{\begin{eqnarray}}
\newcommand{\een}{\end{eqnarray}}

\newcommand{\be}{\begin{equation}}
\newcommand{\ee}{\end{equation}}

\begin{document}

\title{Landau-Zener transition between two levels coupled to continuum }

\author{Rajesh K. Malla}
\affiliation{Center for Nonlinear Studies and Theoretical Division, Los Alamos National Laboratory, Los Alamos, New Mexico 87545, USA}
\author{M. E. Raikh}
\affiliation{Department of Physics and
Astronomy, University of Utah, Salt Lake City, UT 84112}

\begin{abstract}
For a Landau-Zener transition in a two-level system, the probability for a particle, initially in the first level, {\em i}, to survive the transition and to remain in the first level, depends exponentially on the square of the tunnel matrix element between the two levels. This result remains valid when the second level, {\em f}, is broadened  due to e.g. coupling to continuum  [V. M. Akulin and W. P. Schleicht,  Phys. Rev. A {\bf 46}, 4110 (1992)]. If the  level, {\em i}, is also
coupled to continuum, albeit much weaker than the level {\em f}, a particle, upon surviving the transition, will eventually escape. However, for shorter times, the probability to find the particle in the  level {\em i} after crossing {\em f} is {\em enhanced} due to the coupling to continuum. This, as shown in the present paper, is the result of a second-order process, which is an {\em additional coupling between the levels}. The underlying mechanism of this additional coupling is virtual tunneling from {\em i} into continuum followed by tunneling back into {\em f}.

\end{abstract}

\maketitle

\section{Introduction}

The dynamics of a driven quantum system near the avoided level crossing are often exploited by the Landau-Zener dynamics  \cite{Landau,Zener, Majorana, Stukelberg}, where the probability 
of transition into the exciting state is $\exp\left(-\frac{2\pi J^2}{v}  \right)$,
where $v$ is the velocity with which the levels are swept by each other, and $J$ is the tunneling amplitude between the levels at the point of crossing.  In physical systems, the Landau-Zener physics is altered due to coupling to the dissipative environment, which in turn, makes the dynamics non-trivial. In such cases, the transition into the exciting level involves many virtual transitions into the environment. The Landau-Zener dynamics due to the presence of the environment have been addressed in \cite{{Nalbach1,Nalbach2,Nalbach3,Wubs1,Pokrovsky,Javanbakht,Whitney,Huang,Malla1,Malla2,Chen,Ashhab1,Ashhab2,review1}}.

 Here, we investigate the Landau-Zener dynamics when driven two levels are coupled to a continuum. Coupling to the continuum can be effectively described by level broadening. 
 It is nontrivial   that 
{\em for any} $J$ the Landau-Zener formula remains applicable when the second level is broadened \cite{Akulin1992,Vitanov1997}. Even when the inverse width of the final state, i. e. the electron lifetime,
is much shorter than $\frac{J}{v}$, which is
the characteristic time of the transition, the broadening drops out from the survival probability.  
In the latter situation, each time an electron enters the broadened (final) state,
it does not return to the initial state, but rather directly proceeds to the continuum. As a result, at, {\em finite times},
the oscillations in the occupation of the initial state, that take place in the absence of decay,\cite{Vitanov1997}  are washed out due to the
decay. Still, at $t\rightarrow \infty$, the population of the
initial level approaches $\exp\left(-\frac{2\pi J^2}{v}  \right)$ 
asymptotically.

While the result\cite{Akulin1992,Vitanov1997} seems
counter-intuitive, it can be understood in light of
the later developments.\cite{Sinitsyn2004,Shytov}
The key to this understanding lies in the observation that the level broadening caused by the decay can be viewed as a simple smearing of
the level,  {\em f}. On the other hand, the
smearing is equivalent to numerous discrete levels spaced by energy distance much smaller than the width with a matrix element, $J$, being distributed between these levels. For a group
of levels, crossed by the level, {\em i}, the survival probability was  shown to  be equal to the product of the partial probabilities regardless of their separation.\cite{Sinitsyn2004,Shytov} In this way, the broadening, and hence the width, drops out.  

Assume now, that the initial state, $i$, is also decaying albeit much slower than $f$. An
obvious consequence of this decay is that at long times $t\rightarrow \infty$ the particle,
even after surviving the transition, will eventually tunnel out into the continuum. In the present paper we focus on the times much longer than the 
time $\sim \frac{J}{v}$  of the Landau-Zener transition but much shorter than
the decay time of the level {\em i}.  Our prime observation is that,  within this time domain, the combined action of decay of the levels $i$ and $f$ leads
to a strong renormalization of the tunnel matrix element. The most nontrivial outcome of this combined action is that, due to renormalization, the survival probability {\em increases}.

\begin{figure}
	\label{F}
	\includegraphics[scale=0.3]{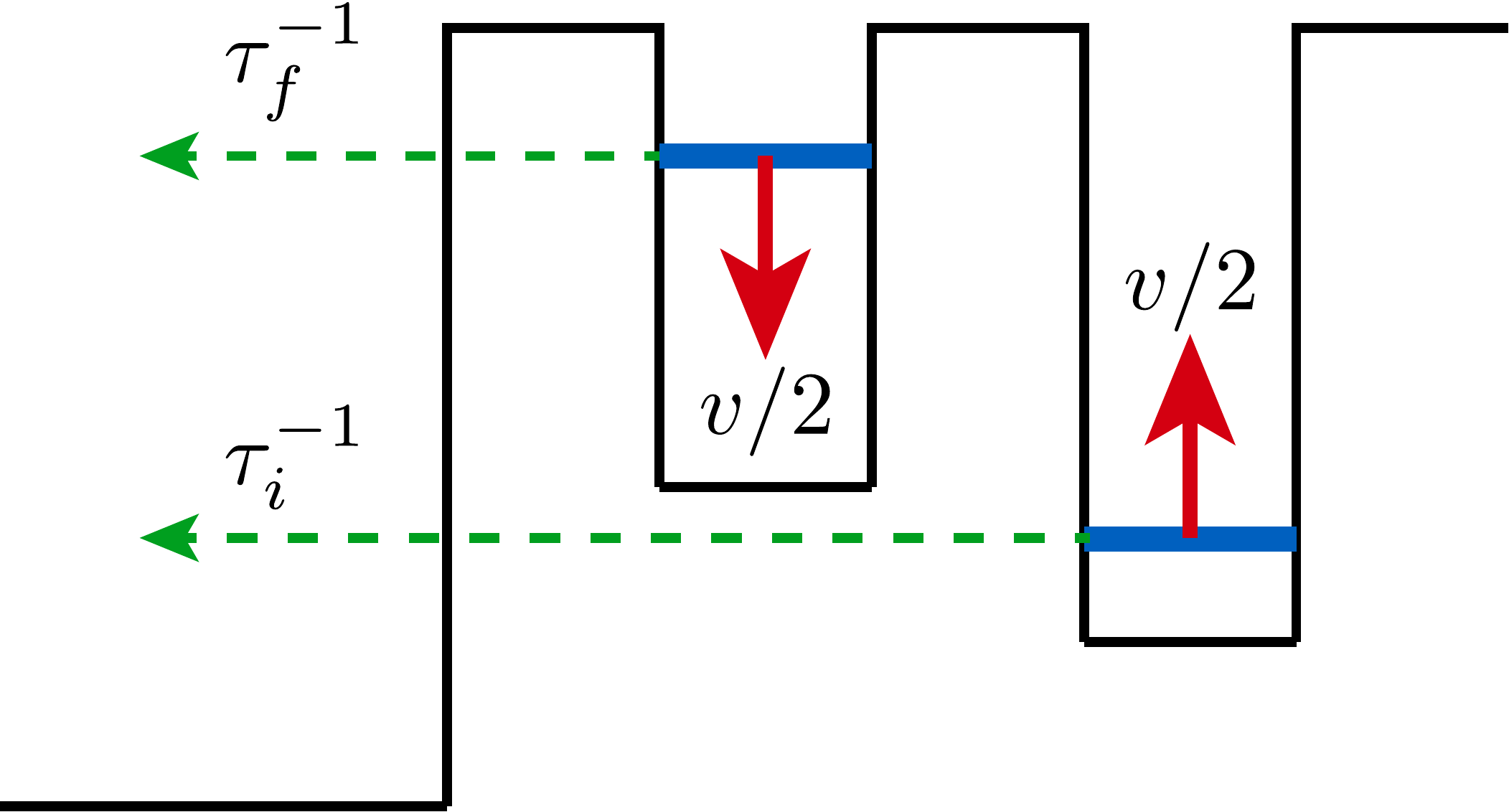}
	\caption{(Color online)
		(a) Illustration of the Landau-Zener transition between two levels, $i$ and $f$, decaying into continuum. As the levels are swept by each other
		with relative velocity, $v$, an electron, initially in $i$,
		can tunnel into the state $f$ and back with the amplitude, $J$. The decay time, $\tau_i$, of the state $i$ is much longer than the decay time, $\tau_f$, of the state,~$f$. }
\end{figure}

\section{\small Coupling of the levels {\em i} and {\em f} via the continuum }

For concreteness we consider a realization of the Landau-Zener 
transition between the two tunnel-coupled quantum dots in which the levels are swept past each other by the gate voltage, see FIG. 1.
As illustrated in the figure, the levels {\em i} and {\em f} are coupled to the {\em same} continuum. The Hamiltonian of the system reads

\begin{align}
\label{Hamiltonian}
&{\hat H}=E_i(t)	C_i^{\dagger}C_i+E_f(t)	C_f^{\dagger}C_f+
\sum_{\nu}\left(V_{\nu i} C_{\nu}^{\dagger}C_i+V_{i \nu}C_i^{\dagger}C_{\nu}    \right)\nonumber\\
&+\sum_{\mu}\left(W_{\mu f} C_{\mu}^{\dagger}C_f+W_{f \mu}C_f^{\dagger}C_{\mu}    \right)+J\left(C_i^{\dagger}C_f+C_f^{\dagger}C_i\right) \nonumber\\
&+\sum_{\nu} E_{\nu}C_{\nu}^{\dagger}C_{\nu},
\end{align} 
where $C_i^{\dagger}$, $C_f^{\dagger}$, and $C_{\nu}^{\dagger}$ are the creation operators of electron in the initial and final states, and the state $\nu$ in the continuum;  $E_i(t)=\frac{vt}{2}$ and $E_f(t)=-\frac{vt}{2}$ are the 
positions of the levels $i$ and $f$, while $E_{\nu}$
is the energy of the state $\nu$ in the continuum. 
Coupling constants $V_{\nu i}$
and $W_{\mu f}$ of the states $i$ and $f$ to continuum are assumed to be
very different, $W_{\mu f}\gg V_{\nu i}$. 

Denote with $\Psi_i({\bf r})$ and $\Psi_f({\bf r})$
the wavefunctions of the initial and final states,
and with $\Psi_{\nu}({\bf r})$ the wavefunction of the
state ${\nu}$ of the continuum. Searching for solution
of the time-dependent Schr{\"o}dinger equation in
the form\cite{Larkin,prigodin}
\begin{equation}
\label{Psi}
\Psi({\bf r},t)=a_i(t)\Psi_i({\bf r})+a_f(t)\Psi_f({\bf r})+\sum_{\nu}a_{\nu}(t)\Psi_{\nu}({\bf r}),	
\end{equation}	
we get the following system of equations for the
amplitudes $a_i$, $a_f$, and $a_{\nu}$

\begin{align} 
\label{System}
&i{\dot a_i}=\frac{vt}{2}a_i+Ja_f +\sum_{\nu}V_{i\nu}a_{\nu},\nonumber\\
&i{\dot a_f}=-\frac{vt}{2}a_f+Ja_i +\sum_{\mu}W_{f\mu}a_{\mu},\nonumber\\
&i{\dot a_{\nu}}=E_{\nu}a_{\nu}+V_{\nu i}a_i
+W_{\nu f}a_f.
\end{align}
 As a next step, we eliminate the states of the continuous
 spectrum from the system. For this purpose,\cite{prigodin} we perform the Laplace transform: 
 ${\overline a=\int_0^{\infty}dt a(t)\exp(-pt)}$
  of the system Eq. \ref{System}. With initial conditions
  $a_i(0)=1$, $a_f(0)=1$, and  $a_{\nu}(0)=0$, we find
\begin{align}
\label{Laplace}	
&i\left(-1+{\overline a_i}\right)=\frac{v}{2}\frac{\partial {\overline a_i}}{\partial p} 
+J{\overline a_f}+\sum_{\nu}V_{i\nu}{\overline a_{\nu}},\nonumber\\
&ip{\overline a_f}=-\frac{v}{2}\frac{\partial {\overline a_f}}{\partial p}+J{\overline a_i}+\sum_{\mu}W_{f\mu}{\overline a_{\mu}}, \nonumber\\
&ip{\overline a_{\nu}}=E_{\nu}{\overline a_{\nu}}+V_{\nu i}{\overline a_i}+W_{\nu f}{\overline a_f}.
\end{align} 
 
 A closed system of equations for ${\overline a_i}$,
 ${\overline a_f}$ emerges upon expressing 
 ${\overline a_{\nu}}$ from the third equation and substituting it into the first and second equations.
 This yields
 
\begin{align}
\label{Clo}
&-i+\left(ip-\sum_{\nu}\frac{V_{i \nu}V_{\nu i}}
	{ip-E_{\nu}} \right){\overline a_i}
-\frac{v}{2}\frac{\partial {\overline a_i}}{\partial p}	\nonumber\\
&=\left(  J+ \sum_{\nu}\frac{V_{i \nu}W_{\nu f}}
{ip-E_{\nu} }  \right){\overline a_f},
\end{align}

\begin{align}	
\label{Clo1}
&\left(ip-\sum_{\mu}\frac{W_{f\mu} W_{\mu f}} {ip-E_{\mu}}   \right){\overline a_f}+\frac{v}{2}\frac{\partial {\overline a_f}}{\partial p}\nonumber\\
&=\left(J+\sum_{\mu} \frac{W_{f\mu} V_{\mu i}} {ip-E_{\mu}}      \right){\overline a_i}.
\end{align}

 When $V_{i\nu}$ and $W_{f\mu}$ depend on $\nu$ and
 $\mu$ only weakly, the sums over $\nu$ and $\mu$
 can be replaced by $\frac{i}{\tau_i}$ and $\frac{i}{\tau_f}$, respectively, where $\tau_i$ and $\tau_f$ are defined as
\begin{align}
	\label{lifetimes}
&	\frac{1}{\tau_i}=2\pi\sum_{\nu}\vert V_{\nu i}\vert^2
	\delta(E_{\nu}-p''), \nonumber\\
	&	\frac{1}{\tau_f}=2\pi\sum_{\mu}\vert W_{\mu f}\vert^2
	\delta(E_{\mu}-p'').
\end{align}
He $p''$ is the imaginary part of $p$. 

The meaning of the times  $\tau_i$ and $\tau_f$
is the decay times from the states $i$ and $f$ 
into the continuum,
respectively. Similarly to Eq. \ref{lifetimes}, the sum in the right-hand sides that adds to
$J$ can be cast in the form
\begin{equation}
\label{Cross}
	\frac{1}{\tau_{if}}=2\pi\sum_{\nu} V_{i\nu}W_{\nu f}
\delta(E_{\nu}- p'').	
	\end{equation}
Inspection of the equations \ref{lifetimes} and
\ref{Cross} leads to an important relation.\cite{shahbazyan} 
 \begin{equation}
	\label{median}	
	\tau_{if}=(\tau_i\tau_f)^{1/2}.
\end{equation} 
This relation suggests that, under the assumption 
$\tau_i \gg \tau_f$,  adopted above, the times $\tau_{if}$ and $\tau_i$ are related as
 $\tau_{if} \gg \tau_i$. Thus, the second-order
process: coupling of the states $i$ and $f$ {\em via the 
continuum} plays a crucial role in the scenario of the
Landau-Zener transition. Namely, in the course of the transition, we can neglect the decay of the state $i$.
After the transition is completed, the survived particle
will eventually escape into the continuum after the 
time~$\tau_i$. 

\section{Survival probability with coupling via the 
	continuum}

With the states of continuum integrated out,
it is convenient to analyze the system of equations
for $a_i$ and $a_f$ in the time domain where it takes
the form
\begin{align}
\label{coupling}
&i\left({\dot a}_i +	\frac{a_i}{\tau_i}\right)=-\frac{vt}{2}a_i+
\left(J+\frac{i}{\tau_{if}}\right)a_f, \nonumber\\
&i\left({\dot a}_f +\frac{a_f}{\tau_f}\right)=\frac{vt}{2}a_f+
\left(J+\frac{i}{\tau_{if}}\right)a_i.
\end{align}

To proceed, we express $a_f$ from the first equation and substitute it into the second. Upon this substitution, we arrive to the closed equation for $a_i$

\begin{widetext}
\begin{equation}
\label{closed}
{\ddot a}_i+{\dot a}_i\left(\frac{1}{\tau_i} +\frac{1}{\tau_f}\right)
+a_i\Biggl\{\frac{iv}{2}+	\Biggl[\frac{vt}{2}+\frac{i}{2}\left(\frac{1}{\tau_i}
-\frac{1}{\tau_f}  \right)  \Biggr]^2+\frac{1}{4}\left(\frac{1}{\tau_i}+\frac{1}{\tau_f}\right)^2 +\left(J+
\frac{i}{\tau_{if}}\right)^2 \Biggr\}=0.
\end{equation}
\end{widetext}
Naturally, in the absence of decay, $\tau_{i}, \tau_{f}, \tau_{if} \rightarrow \infty$, this equation reduces to the conventional equation describing the Landau-Zener transition and yielding   $\exp\left(-\frac{2\pi J^2}{v}  \right)$ for the survival probability. With regard to 
the result of Refs. \onlinecite{Akulin1992}, \onlinecite{Vitanov1997}
setting only two times, $\tau_i$ and $\tau_{if}$, out of three to infinity leaves the result, $\exp\left(-\frac{2\pi J^2}{v}  \right)$, unchanged.
 To trace how this happens, we introduce instead of the amplitude $a_i(t)$ a new variable ${\cal A}_i(t)$ as  follows

\begin{equation}
\label{substitute}
a_i(t)={\cal A}_i(t)\exp\Bigg[-\frac{t}{2}\left(\frac{1}{\tau_i}  
+\frac{1}{\tau_f} \right)\Bigg].
\end{equation}
Substitution Eq. \ref{substitute} allows to eliminate the
first derivative in Eq. \ref{closed} and to reduce it to
the Schr{\"o}dinger-like form. The result of the substitution reads

\begin{align}
\label{Ai}
&{\ddot {\cal A}}_i=\nonumber\\
&-	{\cal A}_i\Biggl\{\frac{iv}{2}+	\Biggl[\frac{vt}{2}+\frac{i}{2}\left(\frac{1}{\tau_i}
-\frac{1}{\tau_f}  \right)  \Biggr]^2 +\left(J+
\frac{i}{\tau_{if}}\right)^2 \Biggr\}.
\end{align}

To get an idea how the short decay  time, $\tau_f$, drops out from the
survival probability, we make use of the fact that, in the long-time limit, the semiclassical
approach applies.
 The semiclassical phase is given by
 
\begin{align}
\label{semiclassic}	
&\Phi(t)=\nonumber\\
&\pm i\int\limits_0^t dt'
\Biggl\{\frac{iv}{2}+	\Biggl[\frac{vt'}{2}+\frac{i}{2}\left(\frac{1}{\tau_i}
-\frac{1}{\tau_f}  \right)  \Biggr]^2 +\left(J+
\frac{i}{\tau_{if}}\right)^2 \Biggr\}^{1/2}.
\end{align}

Keeping only two leading terms, in the long-time limit, we have 
$\Phi(t)\approx \pm i\left[\frac{vt^2}{4} -\frac{t}{2}
\left(\frac{1}{\tau_i} -\frac{1}{\tau_f} \right)\right]$. 
The sign "-" leads to a growing exponent
$\propto \exp\left(\frac{t}{2\tau_f}  \right)$ in ${\cal A_i}(t)$. This term is canceled by 
the prefactor  in Eq.~\ref{substitute}. This cancellation  
illustrates the message of Refs. \onlinecite{Akulin1992},
 \onlinecite{Vitanov1997} about the dropout of the decay. 
 Still, unlike Refs. \onlinecite{Akulin1992},
 \onlinecite{Vitanov1997}, the decay 
 $\propto \exp\left(-\frac{t}{2\tau_i}\right)$ with a long decay time
 survives in $a_i(t)$.

If, at long times, we view Eq. \ref{Ai} as a Schr{\"o}dinger equation,
it can be seen that the corresponding energy is positive  and equal to $J^2$ for $J\tau_{if} \gg 1$, a well-known result for the Landau-Zener transition. However, in the
opposite limit, $J\tau_{if} \ll 1$, which corresponds to the strong  coupling  between $i$ and $f$ via the continuum, the energy
in the effective Schr{\"o}dinger equation is {\em negative} and is equal to $-\frac{1} {\tau_{if}^2}$. Having in mind that the  potential in the effective 
Schr{\"o}dinger equation Eq. \ref{Ai} is the inverted parabola, we conclude that the electron passes the 
transition domain {\em by tunneling}. To trace how
this nontrivial scenario unfolds, we reduce Eq.~\ref{Ai}
to a standard form by introducing a new variable

\begin{equation}
\label{z}	
	z=v^{1/2}e^{\frac{\pi i}{4}}\Biggl[t+\frac{i}{v}\left(\frac{1}{\tau_i} -\frac{1}{\tau_f}\right) \Biggr].
\end{equation}
We see that the new variable 
absorbs the imaginary contribution to time, $t$.
With a new variable, Eq. \ref{Ai} reduces to a standard differential equation for the parabolic cylinder 
functions\cite{Bateman}
\begin{equation}
\label{D}
\frac{d^2{\cal A}_i}{dz^2}+
\left(\nu+\frac{1}{2}-\frac{z^2}{4}  \right){\cal A}_i=0,
\end{equation}
where the parameter $\nu$ is given by
\begin{equation}
\label{nu}
\nu=-\frac{i}{v} \left(J+ \frac{i}{\tau_{if}}\right)^2.
\end{equation}
We see that, as a result of coupling via the continuum,
the parameter $\nu$, describing the transition quantitatively, comes out to be {\em complex}.

Note that for a conventional Landau-Zener transition 
the parameter $\nu$ is given by $\nu=-\frac{iJ^2}{v}$ {\em with a sign minus}. 
To find out how the complex structure  of $\nu$
is reflected in survival probability, we need  
to inspect the solutions of Eq. \ref{D} at complex $\nu$.
To do so, we 
start from the integral
representation of the parabolic cylinder function

\vspace{5mm}

\begin{equation}
	\label{representation}	
	D_{\nu}(z)=\Bigl(\frac{2}{\pi} \Bigr)^{1/2}e^{\frac{z^2}{4}}\int\limits_0^{\infty}	du~
	e^{-\frac{u^2}{2}}u^{\nu}\cos\left(zu-
	\frac{\pi\nu}{2}  \right),
\end{equation}
where the integrals 	$I_\pm(z)$ are defined as
\begin{equation}
	\label{definition}
	I_\pm(z)=	D_{\nu}(z)=\Bigl(\frac{1}{2\pi} \Bigr)^{1/2}e^{\frac{z^2}{4}}\int\limits_0^{\infty}	du~
	u^{\nu}e^{-\frac{u^2}{2}\pm iuz}.
\end{equation}

Following Ref. \onlinecite{Zhuxi}, it is convenient to divide
the integral Eq. (\ref {representation}) into two contributions
\begin{equation}
	\label{two}
	D_{\nu}(z)=	I_{+}(z)e^{\frac{i\pi\nu}{2}} +
	I_{-}(z)e^{-\frac{i\pi\nu}{2}},
\end{equation}
where the integrals 	$I_\pm(z)$ are defined as
\begin{equation}
	\label{definition}
	I_\pm(z)=	D_{\nu}(z)=\Bigl(\frac{1}{2\pi} \Bigr)^{1/2}e^{\frac{z^2}{4}}\int\limits_0^{\infty}	du~
	u^{\nu}e^{-\frac{u^2}{2}\pm iuz},
\end{equation} 
so that $I_{+}(z)=I_{-}(-z)$.  This suggests that the
factors $e^{\frac{i\pi\nu}{2}}$ and $e^{-\frac{i\pi\nu}{2}}$ 
describe the relation of the {\em amplitudes} of the wave function
at negative and positive times. Thus, in calculation of the ratio
of {\em intensities} the imaginary part of $(J+\frac{i}{\tau_{if}})^2$
drops out. This leads us to the final result for the survival probability

\begin{equation}
	\exp\Big(-\frac{2\pi}{v}{\Big\vert} J^2-\frac{1}{\tau_{if}^2}   {\Big\vert}   \Big).
\end{equation}

\section{Concluding remarks}
The prime message of the present manuscript pertains to the papers \onlinecite{Sinitsyn2004}, 
\onlinecite{Shytov}.
According to these papers, the survival probability
is determined by times much longer than the decay time, 
which leads to the conclusion that the
decay of the final state does not affect the survival probability
of the Landau-Zener transition. We point out that the  above
 argument does not apply when the coupling of the initial 
 and the final states via the continuum is taken into account.  Replacing  one final state by the 
 multitude of states with coupling distributed between them emulates the broadening of the final state, and justifies inefficiency of this broadening for
 the survival probability. However, this trick does not
  capture the $\tau_{if}$ process. The ${\tau_{if}}$
  process consisting of
virtual transition of an electron from initial state
into the  continuum followed by virtual transition into the
 final state leads to the formation of the Landau-Zener gap, which is {\em purely imaginary}. Even in the absence of the real gap, $J$, this gap, $\frac{1}{\tau_{if}}$,   determines the
 survival probability. However, after a long time, $\tau_i$,  the electron that survived the transition
 still escapes into the continuum.

Note that the combination $J+\frac{i}{\tau_{if}}$ appeared earlier in Ref. \onlinecite{shahbazyan} in relation to
two-channel resonant tunneling.

It is instructive to compare our results with 
Ref. \onlinecite{Rajesh} where the decay also 
enters into the survival probability. In 
Ref. \onlinecite{Rajesh} the left and right 
localized states in FIG. 1 formed doublets.
The states of the right doublet, emulating the
initial state of the Landau-Zener transition,
did not decay, while both states of the
right doublet, emulating the final state, 
did decay into continuum.
It was shown in Ref. \onlinecite{Rajesh},
that, unlike the conventional transition,
the decay rates of the final states entered 
the survival probability, if they are different. 
By contrast, in the present manuscript we consider
a conventional Landau-Zener transition, but allow 
{\em both} mutually crossing levels to decay.

\vspace{5mm}
\section{Acknowledgement}

R.K.M. would like to thank the U.S. Department of Energy, Office of Science, Basic Energy Sciences,
Materials Sciences and Engineering Division, Condensed Matter Theory Program. R.K.M. was also supported by
the Center for Nonlinear Studies.


\begin{thebibliography}{20}
	\bibitem{Landau} 
	L. D. Landau, ``Zur theorie der energieubertragung,”
	Physics of the Soviet Union {\bf 2}, 46 (1932).
	
	\bibitem{Zener}
	C. Zener,   ``Non-adiabatic crossing of energy levels,” Proc. R. Soc. A {\bf 137} (833), 696 (1932).
	
	
	\bibitem{Majorana} E. Majorana,  ``Atomi orientati in campo magnetico variabile,” Il Nuovo Cimento {\bf 9} (2), 43   (1932).
	
	\bibitem{Stukelberg}  E. C. G. Stueckelberg, “Theory of inelastic collisions between atoms,” Helv. Phys. Acta {\bf 5}, 369  (1932).
	
		\bibitem{Nalbach1}  P. Nalbach, M. Thorwart, ``Landau-Zener Transitions in a Dissipative Environment: Numerically Exact Results," Phys. Rev. Lett. {\bf 103}, 220401 (2009).
	
	\bibitem{Nalbach2} P. Nalbach, M. Thorwart, ``Landau-Zener transitions mediated by an environment: Population transfer and energy dissipation,"
J. Chem. Phys. {\bf 140}, 124709 (2014)
	
	\bibitem{Nalbach3}  P. Nalbach, ``Crossing time in the dissipative Landau–Zener quantum dynamics,"  Eur. Phys. J. B {\bf 95}, 41 (2022).
	
	\bibitem{Wubs1} M. Wubs, K. Saito, S. Kohler, P. H{\"a}nggi, Y. Kayanuma, ``Gauging a Quantum Heat Bath with Dissipative Landau-Zener Transitions," Phys. Rev. Lett. {\bf 97}, 200404 (2006).
	
	\bibitem{Pokrovsky} V.L. Pokrovsky, D. Sun, ``Fast quantum noise in the Landau-Zener transition," Phys. Rev. B {\bf 76}, 024310 (2007)


\bibitem{Javanbakht} S. Javanbakht, P. Nalbach, M. Thorwart, ``Dissipative Landau-Zener quantum dynamics with transversal and longitudinal noise," Phys. Rev. A {\bf 91}, 052103 (2015).
 
 \bibitem{Whitney} R.S. Whitney, M. Clusel, T. Ziman, ``Temperature can enhance coherent oscillations at a Landau-Zener transition," Phys. Rev. Lett. {\bf 107}, 210402 (2011).
 
 \bibitem{Huang} Z. Huang, Y. Zhao, ``Dynamics of dissipative Landau-Zener transitions," Phys. Rev. A {\bf 97}, 013803 (2018).
 
  \bibitem{Malla1} R. K. Malla, E. G. Mishchenko, and M. E. Raikh, ``Suppression of the Landau-Zener transition probability by weak classical noise," Phys. Rev. B {\bf 96}, 075419 (2017).
 
 \bibitem{Malla2} R.K. Malla, M.E. Raikh, ``Landau-Zener transition in a two-level system coupled to a single highly excited oscillator," Phys. Rev. B {\bf 97}, 035428 (2018).
 
 \bibitem{Chen} R. Chen, ``Landau-Zener transitions in a fermionic dissipative environment," Phys. Rev. B {\bf 101}, 125426 (2020).
	
		\bibitem{Ashhab1} S. Ashhab,  ``Landau-Zener transitions in a two-level system coupled to a finite-temperature harmonic oscillator," Phys. Rev. A {\bf 90}, 062120 (2014)
	
	\bibitem{Ashhab2} S. Ashhab, ``Landau-Zener transitions in an open multilevel quantum system
," Phys. Rev. A {\bf 94}, 042109 (2016).

	
	\bibitem{review1} O. V. Ivakhnenko, S. N. Shevchenko, and Franco Nori, ``Quantum Control via Landau-Zener-St{\"u}ckelberg-Majorana Transitions," arXiv:2203.16348 (2022).
	
	\bibitem{Akulin1992} V. M. Akulin and W. P. Schleicht, ``Landau-Zener transition to a decaying level," Phys. Rev. A {\bf 46}, 4110 (1992).
	\bibitem{Vitanov1997}
	N. V. Vitanov and S. Stenholm, ``Pulsed excitation of a transition to a decaying level,"
	Phys. Rev. A {\bf 55}, 2982   (1997).
	
	\bibitem{Sinitsyn2004} N. A. Sinitsyn, ``Counterintuitive transitions in the multistate Landau-Zener problem with linear level crossings,"
	J. Phys. A: Math. Gen. {\bf 37}, 10691 (2004).
	
	\bibitem{Shytov} A. V. Shytov, ``Landau-Zener transitions in a multilevel system. An exact result,"  Phys. Rev. A {\bf 70}, 052708(2004).
	
	
	
	\bibitem{Larkin} 
	A. I. Larkin and K. A. Matveev, ``Current-voltage characteristics of mesoscopic semiconductor contacts," JETP {\bf 66},  580 (1987) [ZhETF {\bf 93}, 1030 (1987)].
	
	\bibitem{prigodin} V. N. Prigodin and M. E. Raikh, ``Decay of the population of quasilocal states in disordered media,"
Phys. Rev. B {\bf 43}, 14073 (1991). 
	
	\bibitem{shahbazyan} 	T. V. Shahbazyan and M. E. Raikh, ``Two-channel resonant tunneling,"
	Phys. Rev. B {\bf 49}, 17123 (1994). 
	
	\bibitem{Bateman} {\em Higher Transcendental Functions}, edited by A. Erdelyi, Vol. 2
	(McGraw-Hill, New York, 1953).
	
	
	\bibitem{Zhuxi} Zhu-Xi Luo and M. E. Raikh, ``Landau-Zener transition driven by slow noise,"
	Phys. Rev. B {\bf 95}, 064305 (2017). 
	
	\bibitem{Rajesh} R. K. Malla and M. E. Raikh, ``Effect of decay of the final states on the probabilities of the Landau-Zener transitions in multistate non-integrable models," 
	arXiv:2204.11782.
	

	
     
\end{thebibliography}

\end{document}